\documentclass{emulateapj}

\shorttitle{Estimating $N$-Point Correlation Functions from Pixelized
  Sky Maps }
\shortauthors{Eriksen et al.}
\usepackage{psfig,epsfig}
\usepackage{amsmath,latexsym}

\usepackage{graphicx}
\usepackage{hhline}
\usepackage{array}
\usepackage{float}
\usepackage{amsmath}
\usepackage{rotating}
\usepackage{subfigure}
\usepackage{epsfig,floatflt}

\begin{document}

\title{Estimating $N$-Point Correlation Functions from Pixelized Sky
Maps} 

\author{H.\ K.\ Eriksen\altaffilmark{1} and P.\ B.\
Lilje\altaffilmark{1}}  

\affil{Institute of Theoretical Astrophysics,
 University of Oslo, P.O.\ Box 1029 Blindern, \\ N-0315 Oslo, Norway}

\email{h.k.k.eriksen@astro.uio.no} 
\email{per.lilje@astro.uio.no}

\altaffiltext{1}{Also at Centre of Mathematics for Applications,
 University of Oslo, P.O.\ Box 1053 Blindern, N-0316 Oslo, Norway}

\author{A.\ J.\ Banday}

\affil{Max-Planck-Institut f\"ur Astrophysik,
Karl-Schwarzschild-Str.\ 1, Postfach 1317,\\ D-85741 Garching bei
M\"unchen, Germany} 
\email{banday@MPA-Garching.MPG.DE}

\and 

\author{K.\ M.\ G\'orski\altaffilmark{2}} 

\affil{JPL, M/S 169/327, 4800 Oak Grove Drive, Pasadena, CA 91109}

\altaffiltext{2}{Also at Warsaw University Observatory, Aleje
 Ujazdowskie 4, 00-478 Warszawa, Poland}
\email{Krzysztof.M.Gorski@jpl.nasa.gov}

\begin{abstract}
 We develop, implement and test a set of algorithms for estimating
 $N$-point correlation functions from pixelized sky maps. These
 algorithms are slow, in the sense that they do not break the
 $\mathcal{O}(N_{\textrm{pix}}^N)$ barrier, and yet, they are fast
 enough for efficient analysis of data sets up to several hundred
 thousand pixels. The typical application of these methods is Monte
 Carlo analysis using several thousand realizations, and therefore we
 organize our programs so that the initialization cost is paid only
 once. The effective cost is then reduced to a few additions per
 pixel multiplet (pair, triplet etc.). Further, the algorithms waste
 no CPU time on computing undesired geometric configurations, and,
 finally, the computations are naturally divided into independent
 parts, allowing for trivial (i.e., optimal) parallelization.

 \end{abstract}


\section{Introduction}

Since the introduction of correlation functions into modern cosmology
by e.g., Totsuji \& Kihara (1969) and Peebles (1973), such functions
have proved to be useful in a large variety of situations. A few
typical applications are the study of the distribution of galaxies and
clusters of galaxies in the universe (e.g., Connolly et al.\ 2002;
Maddox, Efstathiou \& Sutherland 1996; Dalton et al.\ 1994; Croft,
Dalton, \& Efstathiou 1999; Lilje \& Efstathiou 1988; Frieman \&
Gazta\~naga 1999; Jing \& B\"orner 1998), the analysis of the
gravitational lensing shear (e.g., Bernardeau, van Waerbeke, \&
Mellier 2003; Takada \& Jain 2003), or measurements of
non-Gaussianity in the cosmic microwave background (e.g., Eriksen,
Banday, \& G\'orski 2002; Kogut et al.\ 1996).

Unfortunately, higher order $N$-point correlation functions are also
notorious for being computationally expensive. In general the
evaluation of an $N$-point correlation function scales as
$\mathcal{O}(N_{\textrm{pix}}^N)$, and therefore the required CPU time
soon becomes too large to handle.  To remedy this situation several
``fast'' algorithms have been proposed, in particular for analyzing
discrete particle data sets (such as galaxy catalogs). One commonly
used method is to aggregate particles on small scales, and treat each
of these aggregations as a single particle (e.g., Davis et al.\ 1985;
Kaiser 1986). This method effectively corresponds to smoothing the
data set, and is only valid when the distance between the aggregated
particles is much smaller than the scale of interest. It is therefore
of limited use if the primary interest lies in small scales, which
often is the case for CMB data.

The main motivation behind this work is analysis of new
high-resolution CMB maps. In this case, as in several other
applications, the data are not consisting of point sets (e.g.,
positions of galaxies) but in values of a function (like temperature,
polarization, or shear) given for all pixel positions on a map covering
part of or the full sky, where some fixed pixelization scheme has been
used to divide the area of concern into pixels. Further, when
analyzing such maps, the main interest is usually in comparing a
statistic (e.g., a correlation function) estimated on a map of the
real sky with a large Monte Carlo set of the same statistic estimated
on several thousand random realizations based on some theory, e.g.,
Gaussian fluctuations with a given power spectrum. Hence, one must
estimate the correlation function for a set of many thousand similar
maps with the same pixelization to do the analysis of interest.

The $k$d-tree approach (e.g., Moore et al.\ 2001) is based on a
similar idea to the idea of aggregating particles on small scales. It
organizes nearby particles (or pixels) into a hierarchy of bounding
boxes, and uses this hierarchy for rapidly discarding uninteresting
pixel sets. However, it is not obvious that the
$k$d-tree approach is well suited for estimating correlation functions
from a pixelized map (which completely fills the $k$-dimensional
space), unless bin widths much greater than the pixel size is
desired. (Normally one sets the bin width roughly equal to the pixel
size in order to obtain optimal resolution.)

A third approach relies on the Fourier-transform, and this approach
drastically speeds up the algorithms whenever the Fast Fourier
Transform (FFT) is applicable (for a two-point application, see, e.g.,
Szapudi, Prunet, \& Colombi [2001a]). However, this method is not very
attractive for higher-order correlation functions. In the three-point
case one replaces an $\mathcal{O}(N_{\textrm{pix}}^3)$ algorithm with
a very \emph{small} prefactor (multiply three numbers together) with
an $\mathcal{O}(l_{\textrm{max}}^{6})$ algorithm with an extremely
\emph{large} prefactor (compute several Wigner 3-$j$ symbols; see,
e.g., Gangui et al.\ [1994]). Four-point functions are obviously even
more expensive. Thus, the advantage of Fourier-methods for estimating
high-order correlation functions is so far unclear.

In this Paper we describe a new set of algorithms which allows
computation of any (small) subset of the general $N$-point correlation
functions from a pixelized map of the whole celestial sphere, or a
part of the celestial sphere, with up to several hundred thousand
pixels. These algorithms are extremely well suited for Monte Carlo
studies, since we organize the processes so that the very substantial
amount of initialization CPU time is spent only once for the whole
ensemble of sky maps, and not for each individual map. For ensembles
of more than about one thousand realizations, most of the CPU time is
therefore spent on adding pixel values together, not on geometric
computations.

The general idea behind these algorithms is to replace CPU-intensive
inverse cosine operations by much less CPU-intensive
\texttt{if}-tests, at the cost of increasing the memory
requirements. First we compute the distances between any two pixels in
the map and store this information in a set of tables optimized for
fast searches (and compression, if desirable). These tables can be
compared with a set of compasses, in the sense that each column of a
table draws out a circle on the sphere of a given radius, centered on
a given pixel. Then, by searching through two different columns for
equal entries, we generate triangles with the desired size and shape,
and, if necessary, add two or more such triangles to produce $N$-point
multiplets. In fact, our method is equivalent to ruler and compass
construction of triangles.

A similar algorithm for estimating $N$-point correlation functions for
point sets (e.g., galaxies) in three-dimensional space has recently
been developed by Barriga \& Gazta\~naga (2002). However, there are
several important differences between the two methods. First, our
methods are designed particularly for Monte Carlo simulations, in that
we organize the information so that initialization costs are paid only
once. Second, we use the ``compasses'' to determine all three edges in
the triangle, while Barriga \& Gazta\~naga (2002) only use them for
constraining two of the edges. Finally, our methods are particularly
designed to take advantage of the HEALPix\footnote{Available from
http://www.eso.org/science/healpix} nested pixelization scheme
(G\'orski, Hivon, \& Wandelt 1999), although they can (at a
considerable efficiency cost) be generalized to any pixelization.

A first application of a preliminary version of these algorithms to
estimate the three- and four-point correlation functions of the
$COBE$-DMR maps was presented by Eriksen et al.\ (2002), and a
thorough application of the algorithms to the WMAP\footnote{Wilkinson
Microwave Anisotropy Probe} (Bennett et al.\ 2003) data will shortly
be published (Eriksen et al.\ 2004, in preparation). The examples
given at the end of this paper are in fact chosen to correspond with
that analysis.

\section{Overview}
\label{sec:overview}

An $N$-point correlation function is defined as the average product of
$N$ temperatures\footnote{Since our own applications are to CMB data,
especially with the goal of studying non-Gaussianities, we here
concentrate on CMB temperature maps, but $\Delta T$ can of course be
replaced by any scalar quantity.}, as measured in a fixed relative
configuration on the sky,
\begin{equation}
C_{N}(\theta_{1}, \ldots, \theta_{k}) = \biggl<\Delta
T(\hat{n}_{1}) \cdots \Delta T(\hat{n}_{N}) \biggr>.
\end{equation}
Here $\hat{n}_1, \ldots, \hat{n}_{N}$ span an $N$-point polygon with
geometric parameters, $\theta_{1}, \ldots, \theta_{k}$, on the sky,
and $\Delta T(\hat{n})$ is the field value in the direction given by
the unit vector $\hat{n}$.

\begin{figure*}
\plotone{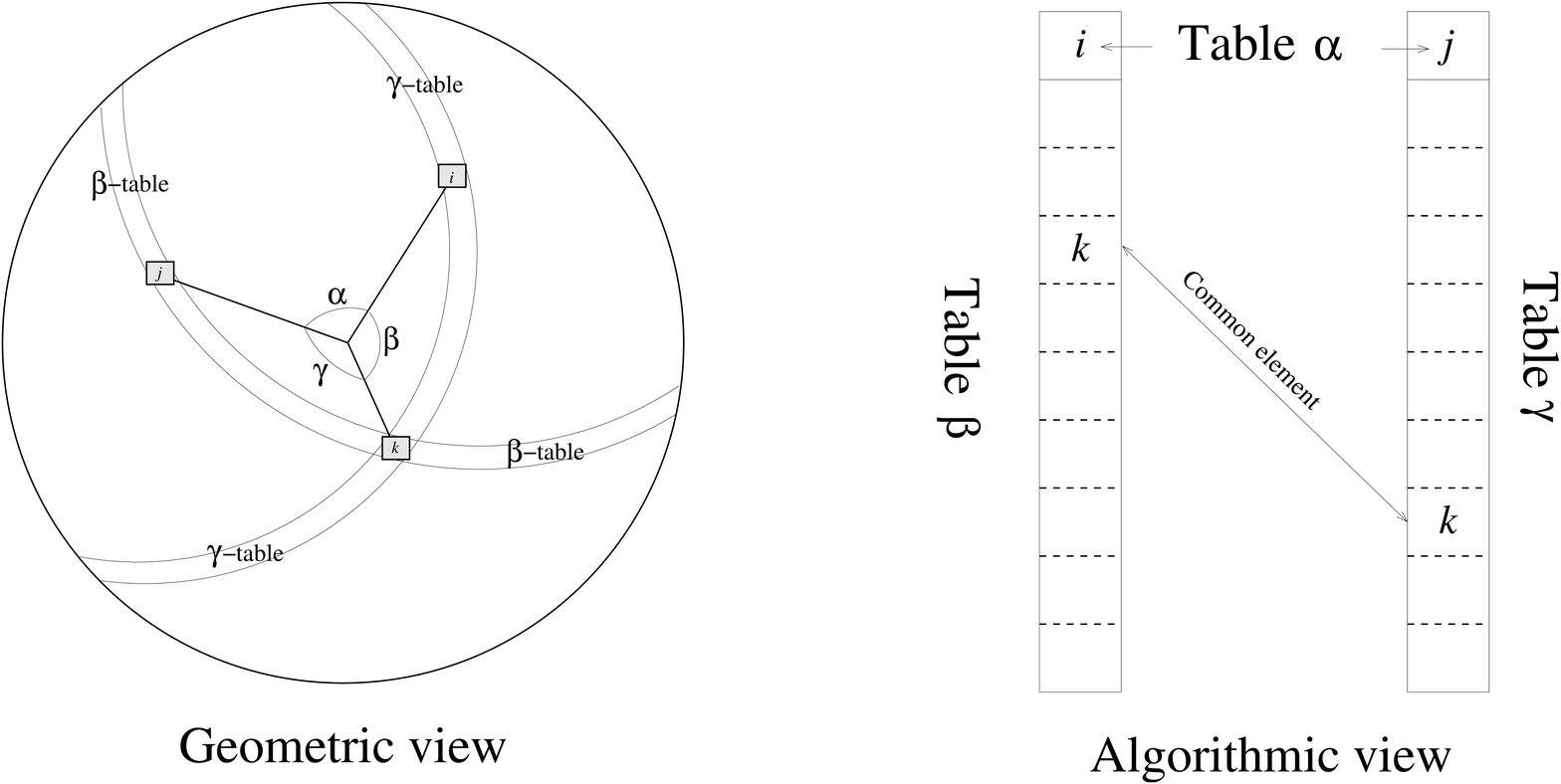}
\caption{Triangles are found by searching through columns of the
  two-point tables pairwise: If we want to find a triangle with
  edges of length $\alpha$, $\beta$ and $\gamma$, we look up a
  pixel pair, $i$ and $j$, in the table corresponding to the angle
  $\alpha$, and let these two pixels define the triangle's base
  line. Next, we look up the $i$'th column of the $\beta$-table, and
  the $j$'th column of the $\gamma$-table. Any common elements in the
  columns will correspond to an acceptable third vertex. Obviously,
  scanning through any such column corresponds to drawing a circle on
  the sky, and the method is therefore equivalent to constructing
  triangles using a pair of compasses.}
\label{fig:gen_triangles}
\end{figure*}

In all common applications we assume isotropy, and then the $N$-point
correlation functions are only dependent on the shape and size of the
$N$-point polygon, and not its particular position or orientation on
the sky. The smallest number of parameters which uniquely determines
such a polygon is $k = 2N-3$.

On a pixelized map, the $N$-point correlation functions are estimated
as simple averages over all pixel multiplets (pairs, triplets,
quadruples etc.)  satisfying the geometric constraints of the
particular configuration,
\begin{equation}
C_{N}(\theta_{1}, \ldots, \theta_{2N-3}) = \frac{\sum_i w_1^i \cdots
  w_N^i \cdot T_1^i \cdots T_N^i}{\sum_i w_1^i \cdots w_N^i}.
\label{eq:def_comp_npt}
\end{equation}
The pixel weights, $w_i$, may be adjusted to account for, e.g., noise
or boundary effects. Note also that we always, but often implicitly,
bin our correlation functions with a given bin width, $\Delta\theta$. A
pixel pair of angular separation $\theta$ is then defined to belong
the the $n$'th bin if $(n-1)\,\Delta\theta \le \theta <
n\,\Delta\theta$.  As seen from equation (\ref{eq:def_comp_npt}), the
evaluation of an $N$-point correlation function is in principle very
simple. However, before we can multiply and add the individual pixel
values, we need to determine all the correct pixel multiplets. That is
less than trivial, and constitutes the main theme of this Paper.

Our starting point is the brute force method for estimating $N$-point
correlation functions. If we want to estimate the three-point
correlation function for only a subset of all possible configurations,
e.g., for all equilateral triangles, the brute force method is to go
through the set of all pixel triplets in the map (resulting in an
$\mathcal{O}[N_{\textrm{pix}}^3]$ algorithm), and compute the three
distances between each pixel pair. Each of the three distances is
computed by taking the inverse cosine of the the dot product of the
two pixel unit vectors. Knowing that one inverse cosine operation
takes about 40 times as many CPU cycles as a multiplication, we see
that not only does the brute force algorithm scale as
$\mathcal{O}(N_{\textrm{pix}}^3)$, but its prefactor is also very
high.  However, the most time consuming shortcoming of the brute-force
approach is that we have to go through every single pixel triplet
existing in the map, even though we may only be interested in a very
small subset of all configurations. Thus, most of the CPU time is
wasted on computing uninteresting information.

Our first task is to produce ``compasses'' for picking out the right
pixels.  Having decided on a number of external parameters, such as
pixel resolution (parameterized by $N_{\textrm{side}}$ of HEALPix),
separation bin width, $\Delta\theta = {\pi}/{N_{\textrm{bin}}}$, and
total number of pixels in the data set, $N_{\textrm{pix}}$ (usually by
introducing a mask), we make ``compasses'' by going through the set of
all pixel pairs, compute the distance between the two pixels in each
pair, and determine which separation bin the pair belongs to. We then
insert the pair into a special table (in the following called a
two-point table) for that bin, a table which is optimized for fast
searches; the $i$'th column of the $k$'th table contains all pixels
with a distance in the interval $[(k-1)\,\Delta\theta,
k\,\Delta\theta)$ from pixel number $i$, sorted according to
increasing pixel number. In other words, the $i$'th column of the
$k$'th table is a ``compass'' that draws out a circle on the HEALPix
sphere with radius $(k-1/2)\,\Delta\theta$ and width $\Delta\theta$,
centered on pixel number $i$.  These two-point tables are computed
once and for all, and stored in individual files for later
use. Already at this point we can estimate the two-point correlation
function with unprecedented efficiency, essentially performing just
one addition for each pixel pair. The particular algorithm for this is
described in \S\ref{sec:calculations}.

However, our main goal is to estimate higher-order $N$-point
correlation functions, and then we have to find all triangles,
quadrilaterals etc.\ on the sky of some given geometric
configuration. Here it is worth noting that once we are able to
construct triangles, all higher-order structures may be generated
simply by putting triangles together. Therefore we here focus on how
to construct triangles.  

As noted previously, construction of triangles is performed by the use
of ``compasses'': suppose that we want to find all triangles with edge
lengths of $\alpha$, $\beta$ and $\gamma$. We then pick out one pair
of pixels, $i$ and $j$, from the two-point $\alpha$-table, and look up
the $i$'th column in the $\beta$-table and the $j$'th column in the
$\gamma$-table. Any common entries in these two tables will have the
correct distances from the two base pixels, and together the three
points span a triangle of the desired size and shape. This search is
fully equivalent to drawing out circles on the sky, and looking for
intersections. The correspondence between the geometric and
algorithmic views is illustrated in Figure \ref{fig:gen_triangles}.

Once this process has been performed for all base pairs in the
$\alpha$-table, we have also found all triangles of the correct shape
and size.  At this point we may therefore forget all about expensive
geometric computations, and concentrate on multiplying and adding the
pixel triplets together as fast as possible.

Let us now go through each step of the underlying algorithms in
detail, paying particular attention to real-world problems, such as
how to deal with limited physical RAM and CPU-time.

\section{Constructing the ``compasses''}
\label{sec:twopt}

The first step in our $N$-point correlation algorithm is to construct
a set of two-point distance tables. Each of these tables
corresponds to a set of ``compasses'' with fixed radius. 


For our idea to be of practical use, the two-point tables must meet
two requirements: first of all, each table must be small enough to
comfortably fit into the physical memory of the computer. Second, the
tables must facilitate fast searches for ``neighboring'' pixels
(defined as pixels within a ring of a given radius and width) for any
given center pixel.

Since we always bin our correlation functions with a given  bin
width $\Delta\theta$, the first requirement is met by
splitting the full $N_{\textrm{pix}} \times N_{\textrm{pix}}$ distance
matrix up according to these bins, simply by letting the bin width be
small enough. Thus, all pixel pairs belonging to the same bin are
collected into one table, which later is stored as a separate file.

The next question is how to organize these tables. The important point
is to be able to efficiently draw out a circle around any given center
pixel. That is, given any position on the sky, we must be able to
easily look up all pixels with a given distance from that position.
For this purpose we choose a simple two-dimensional array, in which
each column corresponds to one single ring on the sky, centered on the
pixel $\mathbf{p}_0$, listed in the zeroth row. Thus, all other
pixels, $\mathbf{p}_j$, in that column satisfies the relation,
$\theta_{\textrm{min}} \le \arccos (\hat{\mathbf{p}}_0 \cdot
\hat{\mathbf{p}}_j) < \theta_{\textrm{max}}$, where
$\hat{\mathbf{p}}_i$ are unit vectors and $\theta_i$ are the angular
limits of the bin.  We will later be searching for identical entries
in two different columns in order to generate triangles, and therefore
we already at this point sort the neighbors of each column according
to pixel number. This facilitates $\mathcal{O}(N_{\textrm{row}})$
searches, as opposed to $\mathcal{O}(N_{\textrm{row}}^2)$ for unsorted
columns.

\begin{figure}
\plotone{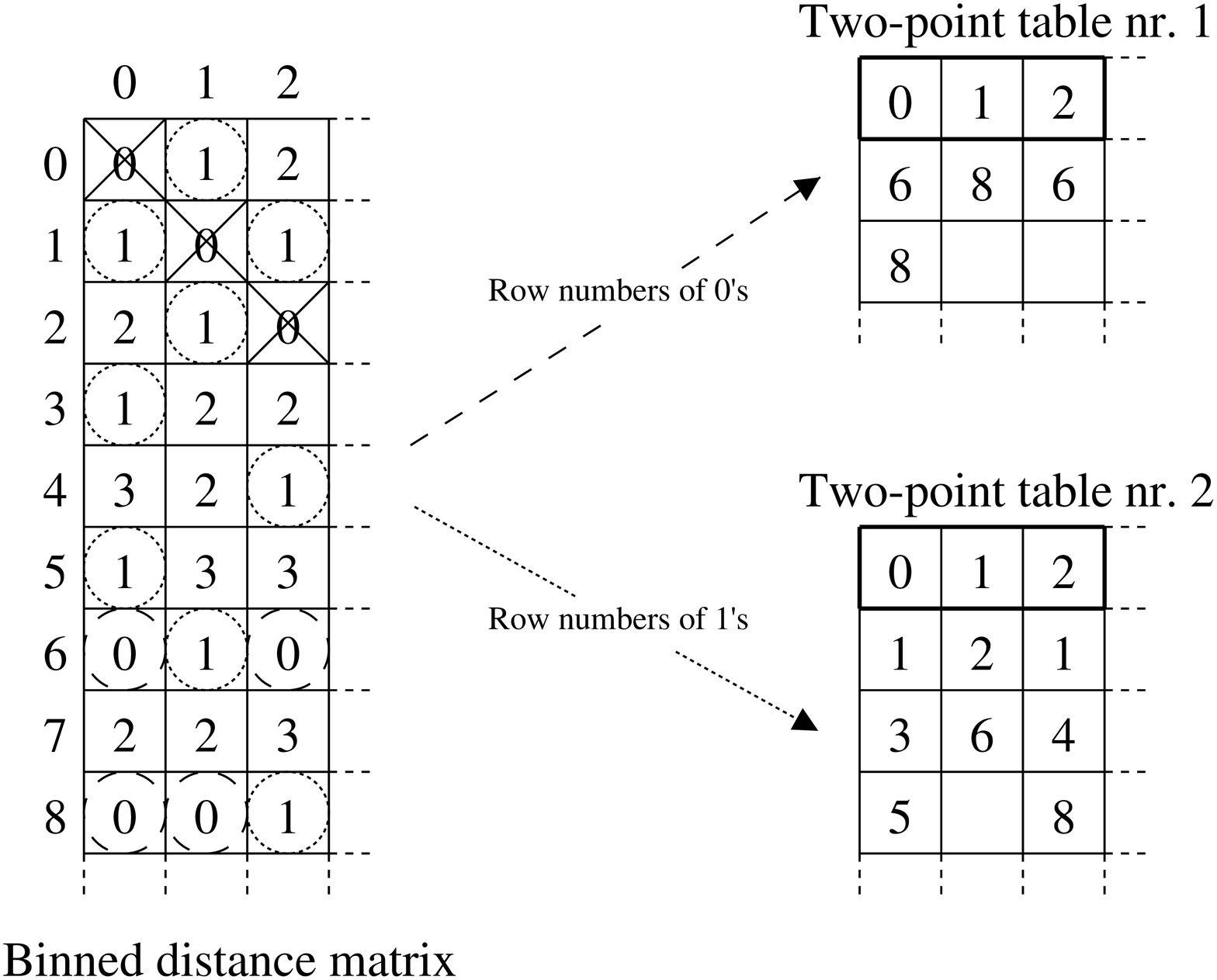}
\caption{The two-point tables are constructed by first computing the
  distances between any two pixels, and then storing the corresponding
  bin numbers in a large $N_{\textrm{pix}} \times N_{\textrm{pix}}$
  array. Then the two-point tables are extracted from this main array
  by collecting all entries with the same binned distance into
  separate tables, maintaining the sorted order of the row (ie.,
  pixel) numbers.}
\label{fig:inmemgen}
\end{figure}

In most experiments we introduce a mask, either in order to exclude
noise-contaminated pixels, or simply to limit our region of
interest. Whatever reason, we incorporate such a mask already at the
construction level, by only including accepted pixels in the
tables. Thus, the zeroth row is nothing but a list of all accepted
pixels, and all pixels in the table are guaranteed to be accepted by
the mask. Not only does this remove many redundant \texttt{if}-tests
later on, but more importantly, it makes sure that the algorithms and
disk usage scale as $\mathcal{O}(N_{\textrm{pix}}^2)$, rather than
$\mathcal{O}(N_{\textrm{side}}^4)$, where $N_{\textrm{pix}}$ is the
number of \emph{accepted} pixels. For small regions in a
high-resolution map, the difference is crucial.

The remaining details are mostly of cosmetic nature. First, since
the two-point tables are simple two-dimensional arrays, there will be
a small amount of unused space in many columns. This extra space is
padded with -1's, distinguishing those entries from valid pixel
numbers. Secondly, we also store all external parameters
($N_{\textrm{side}}$, $N_{\textrm{bin}}$, \texttt{ordering}, bin
number etc.) together with the two-point table in an external file,
making each file an independent entity, in the spirit of
object-oriented programming.


Although the two-point tables are quite simple to describe, they are
surprisingly complicated to generate, the fundamental problem being
their size. Suppose that we have a map with 100\,000 pixels, and we
want to generate a complete set of two-point tables. Doing it
straight-forwardly (i.e., by keeping all information stored in memory
at once), we would need at least 100\,000$^2 \times 4 \,\textrm{bytes}
\approx 10 \,\textrm{GB}$ of memory, and few current computers have
that much physical RAM available. On the other hand, this number is
not prohibitive in terms of disk storage.  The question is therefore
how to generate the two-point tables. Once this operation is
completed, we can store them on a hard drive, compress them if
desirable, and then read only a few at a time when actually estimating
the correlation functions.

We utilize the HEALPix pixelization with \emph{nested} pixel
ordering. This pixelization has the highly desirable property that any
low-resolution pixel may be divided into four high-resolution pixels,
where the pixel numbers of the high-resolution pixels are given by
$4i+k$, $i$ being the pixel number of the low-resolution pixel and
$k=0,\ldots,3$.  

In short the generation process can be described as a three-step
operation:

First we generate a set of tables with low resolution, choosing a
resolution low enough to keep all information simultaneously in the
computer's physical memory.

Second, each table is expanded to a higher resolution by replacing
the low-resolution pixel numbers by the high-resolution pixel
numbers. Since each mother pixel has four children, the output array
is four times as wide and deep as the input array. All pairs in the
output array have distances in approximately the correct range, and
therefore the output two-point table is almost a valid high-resolution
table. However, since the boundary between two rings is smoothed when
the resolution is increased, one may find that a number of pixel pairs
actually belong to the neighboring bins after expansion. This has to
be accounted for, and to do so we sort the pixel pairs in the the
high-resolution output table according to actual bin number, and
output these as incomplete two-point tables.

Finally, after subjecting all low-resolution tables to this operation, we
complete the expansion process by collecting all partial tables of the
same output bin number into complete two-point tables. These
two last expansion steps are repeated as many times as necessary to
obtain the desired resolution.

\begin{figure*}
\plotone{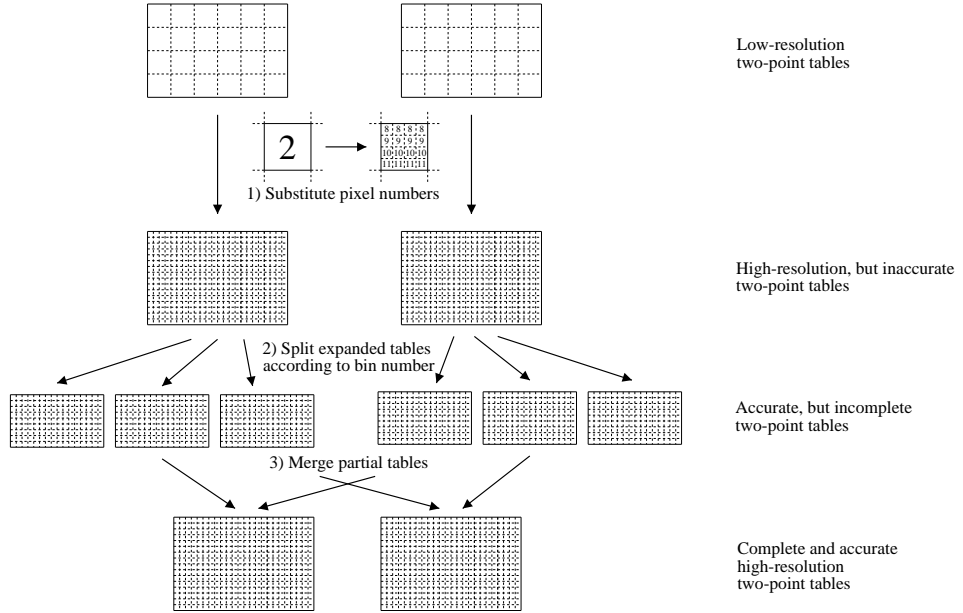}
\caption{Expansion of the two-point tables is carried out in two
  steps. First each entry of the original, low-resolution table is
  substituted with a $4 \times 4$ sub-matrix, containing the pixel
  numbers $4i+k$, where $i$ is the original pixel number and
  $k = 0,\ldots,3$. Then this matrix is split into a set of partial
  two-point tables, according to the correct bin numbers of each pixel
  pair. Finally, the partial two-point tables are merged into
  complete, output two-point tables of higher resolution.}
\label{fig:expansion}
\end{figure*}

In the following three subsections, we describe each of these steps in
more detail.  The first step is illustrated in Figure
\ref{fig:inmemgen} while the two last steps are illustrated in Figure
\ref{fig:expansion}.

\subsection{In-memory generation}

Assume first that we actually do have enough memory available to
simultaneously store all $N_{\textrm{pix}}^2$ pixel pairs in the
computer's RAM, i.e., that there are no memory restrictions. In that
case the procedure is quite simple. First we allocate an
$N_{\textrm{pix}} \times N_{\textrm{pix}}$ array, and run through the
set of all pixel pairs. For each pair we compute the angular distance
$\theta_{ij}$ between the two pixels $i$ and $j$, and mark both the
$j$'th row of the $i$'th column and the $i$'th row of the $j$'th
column by the integer part of $\theta_{ij}/ \Delta\theta$. Once this process
is completed, the desired two-point tables can easily be extracted from
the main array --- the $k$'th table is found by scanning each column
for the number $k$, while storing the row numbers of these entries in
the output array. Note that the output two-point table is
automatically sorted by this method, maintaining the
$\mathcal{O}(N_{\textrm{pix}}^2)$ scaling of the algorithm (rather
than $\mathcal{O}[N_{\textrm{pix}}^2 \, \log N_{\textrm{pix}}]$ as
would be the case if explicit sorting was necessary).

As is indicated in Figure \ref{fig:inmemgen}, we do not count the center
pixel as a valid neighbor to itself in the zeroth bin. Or speaking in
terms of correlation functions, we never let a pixel be multiplied
with itself. For most experiments the noise in each pixel is quite
high, but, fortunately, it is also usually independent from pixel to
pixel. In order to eliminate this noise bias, we exclude powers of
single pixel values from the correlation functions. This is a general
problem for all quadratic estimators.

Let us also make one note on how to compute the binned distance
between two pixels efficiently: the standard method is to take the dot
product of the two pixel center vectors, compute the inverse cosine of
that product, divide the angular distance by the bin width, and keep
the integer part. This method is quite slow because of the inverse
cosine operation which requires about 40 times the CPU time of a
multiplication.  When using the HEALPix nested pixelization, a faster
alternative is available by means of a linear search: one simply
allocates an array, \texttt{binlim}, of length $N_{\textrm{bin}}+1$
which contains the cosine of the bin limits, such that
$\texttt{binlim[i]} = \cos([i-1]d\theta)$ and
$\texttt{binlim[N\_bin+1]} = \cos(N_{\textrm{bin}}d\theta)$.  Then the
correct bin number, \texttt{bin}, may be found by searching for the
array position defined by $\texttt{binlim[bin]} \le \cos\theta_{ij} <
\texttt{binlim[bin+1]}$.

The crucial point here is that the bin position must be stored between
consecutive searches: since the pixels $j$ and $j+1$ are very close to
each other on the sky in the nested HEALPix pixelization, the distance
between $i$ and $j+1$ will be almost the same as the distance between
$i$ and $j$. Therefore one does not have to move more than a few steps
to find the correct bin for each pixel. For the same reason the search
should \emph{not} be implemented by a QuickSearch-algorithm, but
rather with a straight-forward linear method. While the former is
superior for general searches, the latter has considerably less
overhead per step, and is therefore much faster if we only want to
move a few steps, as is the case here.

The above trick is in fact very useful in most two-point correlation
function algorithms, and if the nested HEALPix scheme is employed, one
can often reduce the total CPU time by 60-70\%. Another advantage is
that the bin limits may be set arbitrarily, and therefore the method
seamlessly accommodates for alternative binning schemes. In
particular, Gauss-Legendre binning (ie., placing the bins at the
roots of the Legendre polynomials) is often useful if one wants to
integrate the two-point correlation function to find the power
spectrum (Szapudi et al.\ 1991b).

\subsection{Expanding a two-point table}

The output from the above process is a completely valid set of
two-point tables, but, unless massive amounts of memory is available,
they only describe a low resolution map. A computer with 1GB of RAM
can typically only handle up to 15\,000 pixels through the above
description, and a super-computer with shared memory of $\sim100$GB up
to about 100\,000 pixels. Usually we therefore have to expand the
structure set segment by segment in order to obtain the desired
resolution and pixel number.

As mentioned above, increasing the resolution of a HEALPix map is
particularly simple in the nested pixelization --- each mother pixel is
replaced by four children pixels having pixel numbers given by $4i+k$,
where $i$ is the pixel number of the mother pixel and
$k=0,\ldots,3$. The expansion in resolution is therefore easily
performed by substituting each entry of the original two-point table
by a $4 \times 4$ sub-array (the zeroth row has to be treated
individually, though).  However, since pairs that originate from the
same mother pixel may belong to different high-resolution bins, we
cannot simply substitute the low-resolution pixel numbers by the above
method, and hope that the resulting array will be a valid
high-resolution two-point table.

Usually we also want to increase the number of bins during such an
expansion, considering that a higher resolution supports a smaller bin
width. For these two reasons we have to compute all angular distances
once again to determine to which bins the various pairs belong. Based
on this information we extract a set of partial two-point tables,
similar to what was done in the first, low-resolution step, and store
the resulting partial tables on file, labeled by two numbers,
\texttt{bin\_from} and \texttt{bin\_to}.

\subsection{Merging the partial tables into complete
  high-resolution tables}

The output from the above process is a set of arrays characterized by
the set of parameters \{$N_{\textrm{side}}^{\textrm{from}}$,
$N_{\textrm{side}}^{\textrm{to}}$, $N_{\textrm{bin}}^{\textrm{from}}$,
$N_{\textrm{bin}}^{\textrm{to}}$, \texttt{bin\_from},
\texttt{bin\_to}\}. The final task is to collect all partial tables
with the same \texttt{bin\_to} (containing pixel pairs in the same
output bin, but originating from different low-resolution bins) into
complete two-point tables.  This operation obviously amounts to
merging the partial array columns, while maintaining the already
sorted relationships, and inserting the final result into a large
output array. Since the partial tables are already sorted according to
pixel number, no additional sorting is required. This is actually true
for all the above steps -- no explicit sorting is necessary to
generate sorted output two-point tables.

At this point we have obtained a set of high-resolution two-point
tables stored on disk files. And already at this point we may estimate
the two-point correlation function extremely efficiently. However, our
main goal is to study the higher-order correlation functions, and we
therefore proceed to the construction of triangles by means of the
two-point tables, before turning to applications.

\section{Constructing $N$-point multiplets on the HEALPix sphere}
\label{sec:npoint}

Our algorithm for constructing triangles on the HEALPix sphere is
fully equivalent to ruler and compass construction of triangles (i.e.,
where an equilateral triangle is constructed by first drawing a base
line, then from a given first vertex on this line drawing a circle
with the given length as radius, and then drawing a new circle with
the same radius from the second vertex at the intersection between
the line and the first circle, finding the third vertex of the
triangle at the intersection between the two circles), the only
difference is that the physical compass is replaced with the two-point
tables.

Wanting to find all \emph{equilateral} triangles having a given pixel
$i$ as one vertex and edges given by a given bin number $k$, the first
step is to look up the $i$'th column of the $k$'th two-point
table. All entries in this column are located in the correct distance
from pixel $i$.  Next, we pick an arbitrary pixel, $j$, from this
column. Together these two pixels constitute the base line of the
triangle. Now we scan the $i$'th and the $j$'th column simultaneously,
searching for identical entries. Any common entries in the two columns
will be located in the correct distance from both pixel $i$ and $j$,
and therefore the three pixels together span an equilateral triangle.
Once we find such a triplet we store it in an auxiliary array for
later use. We then keep scanning until the bottom of one column is
reached. Note that since the columns are sorted according to pixel
number, the searches scale as $\mathcal{O}(N_{\textrm{row}})$, rather
than as $\mathcal{O}(N_{\textrm{row}}^2)$ in the unsorted case.

For our ranges of pixel numbers in the maps and bin widths, we
typically find between two and six valid pixels in each such scan. If
we are only interested in scalar valued symmetric three-point
correlation functions, all of these triplets are acceptable, and in
such cases we store all of them for later use. However, if we want to
estimate for instance the four-point correlation function, asymmetric
three-point correlation functions or spin-2 field correlation
functions (for polarization maps), we need to know the orientation of
the triangle, or in other words, we need to know whether we traverse
the triangle clockwise or counter-clockwise when listing the pixels as
\{$i$, $j$, $l$\}. In these cases we introduce the convention that the
two first pixels form the base line and the third pixel is positioned
\emph{above} the base line. In other words, $(\hat{n}_i \times
\hat{n}_j) \cdot \hat{n}_l > 0$.

The above procedure results in a $3 \times n$ array, in which each
column is an acceptable pixel triplet for the current geometric
configuration (in practice we organize the the array slightly
differently, to accommodate for fast summing over the third pixel ---
see \S\ref{sec:calculations} for details). At this point we can
estimate the three-point correlation function by multiplying the three
pixels together, and sum it all up. However, if we want to estimate
the four-point correlation function we need to go one step further.


We find the simplest example of a more complicated geometry when we
want to estimate the rhombic four-point correlation function. This
configuration consists of two equilateral triangles ``glued'' together
on one edge. We can therefore use our list of equilateral triangles to
find the set of all such quadruples. The idea here is simply that the
two triangles must have the same base line, but in reversed order
(since they must have opposite orientation). Thus, we must search for
pairs of triangles for which $i_1 = j_2$ and $j_1 = i_2$, where $i_1$
is the first vertex of the first triangle, $j_1$ is the second vertex
of the first triangle, etc. The four required pixels are then given by
the two common vertices and the two third pixels.

This time we store the sets of four pixels in a $4 \times n$ array,
listed so that the two first entries of each column forms the base
line, the third pixel lies above the base line and the fourth lies
below the base line.

Note that any $N$-point polygon may be built up by triangles, and the
same ideas may therefore be used for computing any $N$-point
correlation function.  In the above examples we have focused on
polygons for which all edges have the same length. However, the same
algorithm can easily be used to also produce general triangles, and
thereby general quadrilaterals. Suppose, e.g., that we want to
find all triangles with edges given by the bin numbers $\{k, l,
m\}$. In that case we would use the $k$'th two-point table to find
pixel pairs, $i$ and $j$, which forms the base line of the
triangle. Then we would simultaneously scan the $i$'th column of the
$l$'th two-point table and the $j$'th column of the $m$'th two-point
table to find the desired third pixels.

\section{Calculating the correlation functions}
\label{sec:calculations}

We start with the case where we want to estimate the two-point
correlation function for separation given by bin number $k$, for which
we already have constructed a two-point table as described above. If
the number of different pairs with separation in this bin is
$N_{\textrm{pair}}$, then the two-point correlation function is
estimated by
\begin{equation}
C_2(k) = \frac{1}{N_{\textrm{pair}}} \sum_{i=0}^{N_{\textrm{pix}}-1}
  \Delta T(\texttt{table}[0,i]) \cdot \sum_{j < i} \Delta
  T(\texttt{table}[j,i]).
\label{eq:comp_twopt}
\end{equation}
Explicitly, we sum up all field values in each column, and multiply
that sum with the zeroth row value. It is difficult to imagine a
more efficient method for estimating the two-point correlation
function than this; we basically perform only one addition for each
pixel pair. Note also that we only sum over pairs for which $i < j$ in
order to avoid double counting, and thus the total CPU time is reduced
by a factor of 2.

For Monte Carlo studies the initialization cost is paid once only, and
in such cases the above equation is where the massive amounts of CPU
time are spent. Although our method intrinsically is quite slow, being
an $\mathcal{O}(N_{\textrm{pix}}^2)$ method, it is for this reason
probably as fast as a two-point correlation function algorithm will
ever be, without introducing either Fourier-methods or large-scale
smoothing.

Let us now take a look at the higher-ordered correlation functions,
and suppose that for any given configuration we have already computed
an $N \times n$ array containing all pixel multiplets satisfying the
binned geometric requirements.  Here $N$ is the order of the
correlation function and $n$ is the number of pixel multiplets. With
such a table the correlation function is estimated by
\begin{equation}
C_N(k) = \frac{1}{n} \sum_{i=1}^{n} \Delta T(\texttt{table}[1,i]) \cdots \Delta
  T(\texttt{table}[N,i]).
\label{eq:comp_npt}
\end{equation}

When the bin width is very small, there is typically only one polygon
corresponding to each base line, and in such cases it is not possible
to accelerate equation (\ref{eq:comp_npt}). However, when the bin
width is comparable to or larger than the pixel size, one often finds
that for two given base line pixels there may be several acceptable
third pixels. If we want to estimate the three-point correlation
function in such cases, we may organize our arrays a little
differently: rather than listing each triplet individually, we may
first list the two base line pixels, then the number of third pixels,
and finally list all those third pixels. Obviously, a one-dimensional
array is more suitable for this organization than the original
two-dimensional one, since the number of third pixels may vary
strongly from set to set.

If the average number of third pixels is larger than about 1.5 (as
often is the case in real-life applications), this linear organization
will both result in a smaller triplet array, and speed up the
computations, since equation (\ref{eq:comp_npt}) is replaced by
\begin{equation}
C_3(k) = \frac{1}{n} \sum_{i=1}^{N_{\textrm{base}}} \Delta
  T_1^i \, \Delta T_2^i \cdot \sum_{j=1}^{N_{\textrm{third}}^i} \Delta
  T_{3,j}^i.
\label{eq:comp_3pt}
\end{equation}
If $\bigl<N_{\textrm{third}}\bigr> = 2$, we here have to perform two
multiplications and two additions for each base pair, a total of four
operations, while in the ``unrolled'' organization, the corresponding
numbers are six multiplications and two additions. Thus the total CPU
time is roughly halved by introducing the linear organization.  The
same trick is even more powerful when estimating four-point
correlation functions, since we in that case can sum over both the
upper and the lower groups of third pixels. Having, e.g., two pixels
in both groups (or in other words, having four different quadruples),
would require twelve multiplications and four additions for each base
pair in the ``unrolled'' organization, but only three multiplications
and three additions in the linear one.

Note also that this linear organization is very simple to
construct, it falls out naturally from the method we use
to construct the triangles in the first place where we fix our interest
on two base line pixels, and scan through the two corresponding
columns for third pixels. Thus, producing the linear organization is
only a matter of storing the results from the scans appropriately.

\section{Additional features}
\label{sec:tricks}

Before benchmarking the algorithms, we briefly discuss a few ideas of
interest to anyone who wants to use these algorithms in practice, and
which are implemented in our software. First we show how to optimize
the cache usage in the case of Monte Carlo studies, and then we
briefly comment on how to parallelize these algorithms. Finally, we
show how to compress the two-point tables in order to save disk space.

\subsection{Mask-based pixel numbering and cache usage}

From the previous sections it should be clear that these $N$-point
algorithms are very close to being optimal for Monte Carlo
simulations. In applications where the initialization cost is small
compared to the computational costs, there is not much left to gain
from an algorithmic point of view --- we have removed all geometry from
the computations, and all that is left is to multiply field values
together as fast as possible, and sum it all up. However, there are
two main issues left to consider, cache optimization and
parallelization. Let us start by looking at the cache usage.

In a Monte Carlo application, we want to estimate the two-point
correlation function for, say, 1000 different maps. The naive approach
is simply to loop over the maps, applying equation
(\ref{eq:comp_twopt}) repeatedly. However, this has two negative
effects. First of all, the CPU has to move the two-point table
information from the main memory into the cache 1000 times. Secondly,
the number of hits per memory page in the map array may be less than
optimal, since consecutive rows may contain widely separated pixel
numbers. On the other hand, both the fact that the columns of the
two-point tables are sorted, and the nested HEALPix pixelization, 
improve the situation.

If we have considerable amounts of physical memory available, we can
do better by analyzing all maps simultaneously. In this case we store
all 1000 maps row-wise in a two-dimensional array, so that each column
contains the map values of one single pixel. With this large
super-array we can loop over the maps in the innermost loop, rather
than in the outermost. Thus we are guaranteed almost perfect cache
usage. Also, we only need to look up the two-point table once, further
reducing the memory traffic.

The question is then, do we have enough memory available to store all
1000 maps? For a map of, say, three million pixels, the answer is
clearly ``no'', since that would require about 12 GB of memory,
assuming single precision floating point numbers. On the other hand,
for maps with less than 50\,000 pixels we only need 400 MB, and that
is not a large amount for current computers. 

Even with our algorithms, estimation of $N$-point correlation
functions is CPU intensive, and maps of several million pixels are out
of reach for today's computers. Thus, CPU time is the limiting factor,
rather than available memory. For this reason we often choose to
analyze high-resolution maps region by region (by the introduction of
masks), where each region contains no more than, e.g., 50\,000
pixels\footnote{This method was recently used by \citet{Eriksen:2004}
in an analysis of the \emph{WMAP} data. The largest scales were probed
by first degrading the maps from $N_{\textrm{side}}=512$ to
$N_{\textrm{side}}=64$, and then computing each correlation function
in $1^{\circ}$ bins on a set of complementary hemispheres. The smaller
scales were studied by partitioning the sky into $10^{\circ}$ disks,
on each of which the three-point function was estimated.}. In these
cases we can discard all the unused map information, and retain the
included pixels only.

Explicitly, we now introduce a new pixel ordering scheme, counting
relative to the mask rather than to the resolution parameter
$N_{\textrm{side}}$. The maps are converted from the standard HEALPix
numbering to the mask-based ordering by extracting all included
pixels, while conserving their relative order. Thus, the first
included pixel is assigned the new pixel number 1, the second included
pixel is assigned the new pixel number 2 etc. By storing these
conversion rules in an external table, we may easily go back and forth
between mask-based and map-based pixel ordering.  In order to optimize
the cache usage for high-resolution maps, we convert all maps and
two-point tables to mask-based pixel ordering, and then proceed as
before.

\subsection{Parallelization of computational routines}

The final issue to consider from an efficiency point of view is
parallelization. There are two different cases, the geometric
computations and the actual correlation function calculations.

The latter part is quite trivial, considering that we already in the
algorithmic design have divided the problem into completely separate
parts --- usually we want to compute many bins of each correlation
function, and then we let each processor work on its own
geometric configuration. This obviously results in perfect speed-up
(i.e., doubling the number of processors halves the total CPU time).

Parallelization of the construction of the two-point tables is not
quite as simple, at least not for the very first step, in which we
generate a full $N_{\textrm{pix}} \times N_{\textrm{pix}}$
table. However, one possibility is to let each processor work on
separate columns, and then merge the partial results into one complete
array in the end. Once this task is performed, the problem is once
again divided into separate parts, and the expansion steps may be
trivially parallelized, like the $N$-point calculations.

\subsection{Compression of the two-point tables}

If disk space is a limiting factor, one may be discouraged by the
substantial size of the two-point tables. In such cases one may wish
to implement file compression.  One common algorithm for this purpose
is the Huffman (1952) coding method. The idea behind this algorithm is
to assign short bit strings to frequently occurring symbols (i.e.,
integers in our setting), and long bit strings to rarely occurring
symbols. The average number of bits per symbol is thus minimized.
Applied to our two-point tables, it appears at first sight as though
little is gained; since any pixel number has to occur somewhere in
each column, each number is more or less equally frequent in each
two-point table. However, one property of our two-point tables still
makes Huffman coding most efficient, namely the fact that each column
is sorted according to increasing pixel number. Therefore, if we
subtract the values in two consecutive rows, we get a small positive
number, and the total histogram of the table is strongly skewed toward
low values. This new, transformed array may be compressed very
efficiently.

A simple way of taking advantage of this idea is to do the following:
first we transform the original array by subtracting consecutive rows,
and store the resulting array in an external file. Then we compress
that file using any available compression utility (which usually
relies on a combination of Huffman and run-length coding), e.g.,
\texttt{gzip}. This method routinely reduces the required disk space
by 80-90\% for standard combinations of $N_{\textrm{side}}$ and
$N_{\textrm{bin}}$. 

In the current implementation there are special routines for reading
and writing two-point tables. Compression may therefore be enabled
simply by adding a few appropriate lines in each of those routines
(subtracting consecutive rows, making a system call to
\texttt{gunzip}/\texttt{gzip} etc.), leaving the rest of the source
code unchanged.

\section{Benchmarking the algorithms}
\label{sec:benchmarking}

We have run several tests to determine what resources are required for
application of the above algorithms. The parameters of these tests
mimic those used in the small-scale WMAP analysis of
\citet{Eriksen:2004}, in that we let our region of interest be a disk
on the HEALPix sky, pixelized at $N_{\textrm{side}} = 512$, and with a
bin width of $7\farcm 2$. The only difference is that we let the
radius of the disk vary between $5^{\circ}$ and $15^{\circ}$ (in steps
of $2\fdg 5$) in order to see how the algorithms scale with respect to
number of pixels. These disks correspond to pixel numbers between 5940
and 53\,464, and we should therefore get a good picture of how the
algorithms behave in real-world situations.  The computer used in the
following tests is a Compaq EV6 Alpha server.

We start by generating a set of two-point tables for each disk,
and measure the CPU time and disk space spent in this process. Next we
estimate the two-, three- and four-point correlation functions from
100 simulated maps, and find the number of pixel multiplets (pairs,
triplets and quadruples), the CPU time per realization and
initialization cost as a function of $N_{\textrm{pix}}$.

\subsection{Wall-clock time and disk space for generating the
  two-point tables}

\begin{figure*}
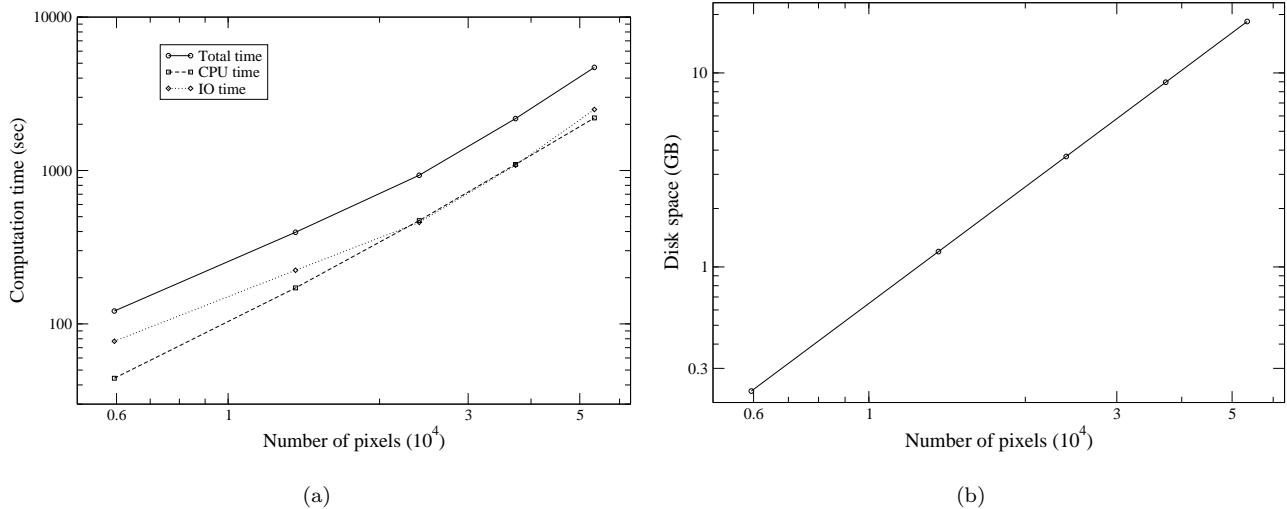

\mbox{\subfigure[]{\label{fig:struct_gen_cputime}\epsfig{figure=f4.eps,width=0.46\textwidth,clip=}}
      \quad
      \subfigure[]{\label{fig:struct_gen_diskspace}\epsfig{figure=f5.eps,width=0.46\textwidth,clip=}}}
\caption{a) The time required for generating a set of two-point tables
for disks of varying size. Note that the I/O time (i.e., disk
activity) is of the same magnitude as the CPU time, reflecting the
massive amounts of data that are processed in these
operations. Neither of these functions follows a power law. b) The
disk space requirements for storing the two-point tables. This
function follows a nearly perfect $\mathcal{O}(N_{\textrm{pix}}^{2})$
relation, with an effective index of 1.99.}
\label{fig:struct_gen}
\end{figure*}

The results from the two-point table generation tests are shown in
Figure \ref{fig:struct_gen}. In panel (a) we see that the time spent
on this process does not follow the expected
$\mathcal{O}(N_{\textrm{pix}}^2)$ relationship. In fact, the CPU time
scales as $\mathcal{O}(N_{\textrm{pix}}^{1.78})$, while the
input/output time (i.e., disk activity) doesn't even follow a power
law. The explanation is the same in both cases: each of these
quantities follows a relation of the general form, $T_i = a_i + b_i
N_{\textrm{pix}} + c_i N_{\textrm{pix}}^2$. Algorithmic overhead and
initialization costs are typically linear in $N_{\textrm{pix}}$, and
these are likely to dominate for a small number of pixels. In other
words, $b \gg c$. What we see in Figure \ref{fig:struct_gen_cputime} is
the manifestation of this. For the relatively modest number of pixels
we consider, we do not reach the asymptotic $N_{\textrm{pix}}^2$
region, and therefore the effective scaling coefficient in this region
is smaller than 2.

This result may seem somewhat counter-intuitive; normally a low
scaling coefficient is interpreted in a positive direction. That is
not the case here. Rather, it may suggest that the current
implementation is sub-optimal, and that there is room for
improvements. However, it should also be noted that a truly
$\mathcal{O}(N_{\textrm{pix}}^{2})$ implementation can probably only
be devised if one is able to store the full $N_{\textrm{pix}} \times
N_{\textrm{pix}}$ distance matrix in the physical RAM simultaneously,
and in that case our methods are superfluous.

The important conclusion, however, is that the current algorithms are
in fact sufficiently efficient to handle data sets with at least a few
hundred thousand pixels, considering that it only takes about an hour
to generate a set of two-point tables for 50\,000 pixels.

In Figure \ref{fig:struct_gen_diskspace} the corresponding disk space
usage is shown. Here we do see a virtually perfect
$\mathcal{O}(N_{\textrm{pix}}^{2})$ scaling, a result which should
come as no surprise. The linear overhead for storing the two-point
tables is basically just a few extra rows (for storing the pixel
centers and padding extra space with -1's), and this is normally
negligible compared to the total number of ``neighboring'' pixels.

Note also that disk space is not a critical factor in these
computations; a data set of 50\,000 pixels requires less than 20 GB of
space, a trivial amount of disk space for current computers. If we
want to include as many pixels as 200\,000, we would need 160 GB, and
then perhaps compression might be needed, as described in the previous
section. Then the disk space requirements would be less than 30 GB,
and once again well within the limits of current computers.

We may therefore conclude that the first step in the two-point table
strategy, the construction of the two-point tables, do not pose a
serious problem neither in terms of wall-clock time nor disk space,
when analyzing data sets which contain fewer than a few hundred
thousand pixels.




\subsection{Estimating correlation functions}

\begin{figure*}
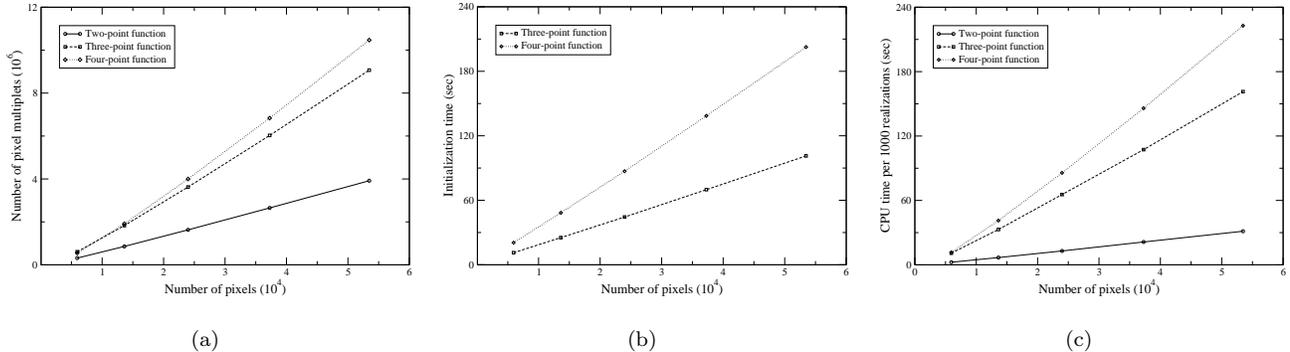

\mbox{\subfigure[]{\label{fig:corr_num_multi}\epsfig{figure=f6.eps,width=0.3\textwidth,clip=}}
      \quad
      \subfigure[]{\label{fig:corr_init}\epsfig{figure=f7.eps,width=0.3\textwidth,clip=}}
      \quad
      \subfigure[]{\label{fig:corr_comp}\epsfig{figure=f8.eps,width=0.3\textwidth,clip=}}}
\caption{a) The number of pixel multiplets as a function of
  $N_{\textrm{pix}}$ for the three different function types. Note that
  these numbers scale almost linearly with $N_{\textrm{pix}}$,
  boundary effects make up the difference. b) The CPU time required
  for generating the set of equilaterals and rhombi, and c) the CPU
  time for computing one configuration of each $N$-point correlation
  function for 1000 realizations.}
\label{fig:comp_resources}
\end{figure*}

Finally, we check that the amount of time spent on actually computing
the correlation functions is not prohibitively large. This experiment
is carried out by measuring both the initialization and the
addition/multiplication time for each of the two-, three- and
four-point correlation functions, for one particular configuration on
each of the disks described above. The configurations were arbitrarily
chosen to be the $3^{\circ}$ two-point, equilateral three-point, and
rhombic four-point configurations.

The results are shown in Figure \ref{fig:comp_resources}. In
panel (a) we see the numbers of pairs, triplets and quadruples
contributing to the bin as a function of $N_{\textrm{pix}}$. Note that
these numbers increase almost linearly with the number of pixels in the
data set, with best-fit power-law indexes of 1.15, 1.23 and 1.34,
respectively. This is an intuitive result; doubling the
area of interest also doubles the number of polygons that fit into
it. The deviation from a perfect linear relationship is caused by
boundary effects.

Given this linear relationship, we also expect the CPU time required
for multiplying and adding the pixel multiplets together to increase
linearly with $N_{\textrm{pix}}$. In Figure \ref{fig:corr_init} we see
that this is the case. It is important to note that the CPU time for
computing any single bin of any $N$-point correlation function
increases \emph{linearly} with the number of pixels in the data set.

It is well known that the evaluation of a general $N$-point
correlation function scales as
$\mathcal{O}(N_{\textrm{pix}}^{N})$. However, this relation only
applies to the computation of the full correlation function, i.e.,
when including all $N_{\textrm{pix}}^{N}$ pixel multiplets with all
possible multiples of the bin size as lengths of the edges. In most
applications, this is found unfeasible, and one uses only a small
subset of all available geometric configurations of the $N$
vertices. In those cases the computational scalings are less severe;
the cost for computing one single bin increases only proportionally with
the number of pixels in the data set, and that time is not radically
different for the two-, three-, and four-point correlation
functions. Only the number of different available configurations
change from correlation function to correlation function, not the cost
for computing one given separation bin.

In Figure \ref{fig:corr_init} and \ref{fig:corr_comp} we see that for
a data set with 50\,000 pixels it takes about 90 seconds to find all
equilateral triangles with $3^{\circ}$ edges, and 200 seconds to find
all rhombi. In general it takes about twice as long to find all
quadruples as finding all triangles, an obviously logical result ---
quadrilaterals are constructed by merging two triangles. However, it
is worth noting that this extra cost can be avoided when computing
quadrilaterals constructed from isosceles triangles, since one then
does not have to worry about the orientation of the triangles.

As mentioned in the introduction, our methods are designed for Monte
Carlo studies, and in Figure \ref{fig:corr_comp} the CPU time for
analyzing 1000 realizations simultaneously is shown, excluding
initialization. We see that for a 50\,000 pixel data set it takes about
30 seconds to estimate one two-point function bin, 150 seconds
for one three-point configuration and 210 seconds for one four-point
configuration. For the higher-order correlation functions these
numbers are about the same as the initialization costs, and we
therefore conclude that it is crucial to organize our programs so that
the initialization costs are paid once only.

A typical example of an $N$-point correlation function analysis could
be the following: We analyze 5000 Monte Carlo realizations on a disk
containing about 50\,000 pixels, with 100 configurations (distance
bins) in the two-point correlation function, 500 different
configurations in the three-point correlation function, and 1000
different configurations in the four-point correlation function. We
assume that the time for computing each of these configurations is
constant and equal to the $3^{\circ}$ configurations. While this
obviously is not strictly true (more pixel multiplets are associated
with larger configurations), this particular size is a good
average of what is used in real-world CMB analysis.

First we have to generate the two-point tables for the disk, a process
which is performed only once. The total wall-clock time for this is
about 3 hours, and we need about 20 GB of disk space to store the
tables. Next, about 1 GB of RAM is required for storing the maps, and
an additional few hundred MB for generating the triangles and
quadrilaterals. This amount of disk space and RAM is not a major
obstacle for current workstations or super-computers.

Next, the estimation of the two-point correlation functions takes about $100
\cdot 5 \cdot 30 \,\textrm{sec} \approx 5 \,\textrm{hours}$, while the
three-point configurations require $500 \cdot (150 \,\textrm{sec} + 5
\cdot 150 \,\textrm{sec}) \approx 125 \,\textrm{hours}$. Finally, the
four-point configurations need $1000 \cdot (200 \,\textrm{sec} + 5
\cdot 220 \,\textrm{sec}) \approx 400 \,\textrm{hours}$. None of these
numbers are prohibitively large, and we can therefore conclude that
the proposed algorithms are able to handle real-world data sets using
reasonable computational resources. 

\section{Conclusions}

We have described a set of algorithms for estimating $N$-point
correlation functions from a pixelized map. The fundamental idea is
first to locate all pixel multiplets (pairs, triplets, quadruples
etc.) corresponding to one given geometric configuration, store these
in a table, and then finally use that information to efficiently add
and multiply the pixel values together. The pixel multiplets are found
by searching through a set of two-point tables, motivated by ruler and
compass construction of triangles.

The construction of the two-point tables only take a small amount of
CPU time (typically a few hours), and since this operation is performed
only once, it is completely negligible compared to the following
operations. The storage requirements of these tables are well within
the limits of current computers. 

We have found that the initialization cost for each configuration
roughly equals the computation time of 1000 realizations, so for a
Monte Carlo simulation with, say, 5000 realizations, the major amount
of time is spent on adding pixel values together. Thus, the geometric
computations are removed from the problem. In other words, our methods
are probably as fast as any fully $\mathcal{O}(N_{\textrm{pix}}^N)$
algorithm can ever be, when applied to a Monte Carlo analysis.

However, the main advantage of our method over a brute-force analysis
is the fact that we estimate the correlation functions
configuration-by-configuration -- no CPU time is spent on
uninteresting configurations. This effectively reduces the scaling
of an experiment in which only a fixed number of configurations is
desired (such as only equilateral triangles, rhombic quadrilaterals, or
configurations covering scales smaller than, say, $2^{\circ}$) to
$\mathcal{O}(N_{\textrm{pix}})$. Thus, even large data sets may be
subjected to an $N$-point correlation function analysis.

Another consequence of this configuration-wise division is optimal
parallelization. By letting each processor work on separate
configurations, optimal speed-up is obtained, at the cost of some extra
RAM requirements.


\acknowledgements

HKE and PBL thanks the Research Council of Norway for economic
support, including a PhD studentship for HKE. This work has also
received support from The Research Council of Norway (Programme for
Supercomputing) through a grant of computing time. The authors
acknowledge use of the HEALPix (G\'orski et al.\ 1999) software and
analysis package for deriving the results in this paper.

\end{document}